# Topological interface states induced by incident angle in the 1D elastic wave system

Pan Li[1], Wenping Hu[1], Pai Peng[2], Xuefeng Zhu[1], Degang Zhao[1†]

[1]School of Physics, Huazhong University of Science and Technology, Wuhan 430074, China
[2] School of Mathematics and Physics, China University of Geosciences, Wuhan 430074, China
† dgzhao@hust.edu.cn

Topological interface states are currently attracting rapidly growing attention in classical wave systems. However, little work has been done on topological interface states in one-dimensional (1D) elastic wave systems, especially in the case of oblique incidence. This paper theoretically demonstrates the realization of topological interface states of elastic waves in a 1D composite plate structure composed of two phononic crystals (PCs) with different topological characteristics, which can be regulated by the incident angle. For the out-of-plane SH mode, multiple topological interface states can coexist in different common bandgaps. For the in-plane complex P-SV coupled mode, topological interface states can exist in both "partial-polarization" and "omni-polarization" bandgaps. All these interface states are in the wide frequency and incident angle regions. We also discuss the polarization and the mode conversion of the interface states. Our results provide an innovative method to excite and tune topologically protected interface state for elastic waves, which may have potentially applications in obtaining strong local vibration for different polarized elastic wave modes.

## I. Introduction

In recent years, the concept of topology in the field of condensed matter has led to a series of new physics, such as the quantum anomalous Hall effect [1], Weyl semimetals [2], and topological insulators [3]. The most exotic and attractive feature of topological systems is the topologically protected interface states [4-8]. As they are robust against local perturbations and disorders, topological interface states have attracted a lot of attention in the fields of optics [6], acoustics [7], and mechanics [8].

In the one-dimensional (1D) periodic system, the topological properties of the bulk band structure can be characterized by the Zak phase [9]. The concept of the Zak phase was firstly proposed in a condensed matter system and later extended to optical and acoustic systems. Xiao et al. firstly found the general connection between the existence of the topological interface states and the Zak phase in a photonic system [10]. Soon after, they theoretically and experimentally demonstrated the acoustic topological interface states in a system composed of cylindrical waveguides [11]. After that, different structures, such as conventional photonic crystal [12,13], chiral photonic crystal [14], and dielectric resonator chain [15], were used to obtained optic topological interface states. However, most have been done in optical systems was about normal incidence. Recently, Hu et al. studied the case of oblique incidence and achieved spin-dependent secondary topological interface states in optical complex superlattices [16]. Like optical systems, the topological properties of 1D acoustic systems also attracted a lot of attention. Some artificial structures were proposed to achieve acoustic topological interface states in the Bragg bandgaps [17-20] or in the local resonant bandgaps [21-25]. In elastic wave systems, the topological interface states of the transverse waves [26-29] and longitudinal waves [30,31] were reported sequentially. Nevertheless, most of them only considered the case of normal incidence, in which the longitudinal waves and transverse waves can be decoupled. The case of oblique incidence in elastic wave system remains to be further studied.

This article systematically demonstrates the topological interface states in the case of oblique incidence in a 1D elastic wave system. According to the polarization pattern, elastic waves can be divided into longitudinal waves (so-called P waves) and transverse waves (so-called S waves). For transverse waves, those whose vibration direction is perpendicular to the incident surface are so-called SH waves, and those whose vibration direction is parallel to the incident surface are so-called SV waves. For the oblique incidence, the SH mode is an out-of-plane mode, which can be separated individually. While both the P mode and SV mode are the in-plane modes, and they will interact to form a hybrid mode. The reflection and refraction styles of SH mode and P-SV coupled mode are schematically shown in Figs. 1(a) and 1(b), respectively. For these two cases, we systematically study the evolution of band structures as well as topological phases by the transfer matrix method [32,33]. Results show that topological properties of bulk bands can be tuned by the incident angle. For the SH waves, multiple topological interface states are produced on the interface of one composite structure which is composed of two phononic crystals (PCs) with different topological phases. For the P-SV coupled mode, we achieve topological interface states in both "partial-polarization" and "omni-polarization" bandgaps. Our work has potential significance to modulate the mode polarization of topological interface states in elastic wave system.

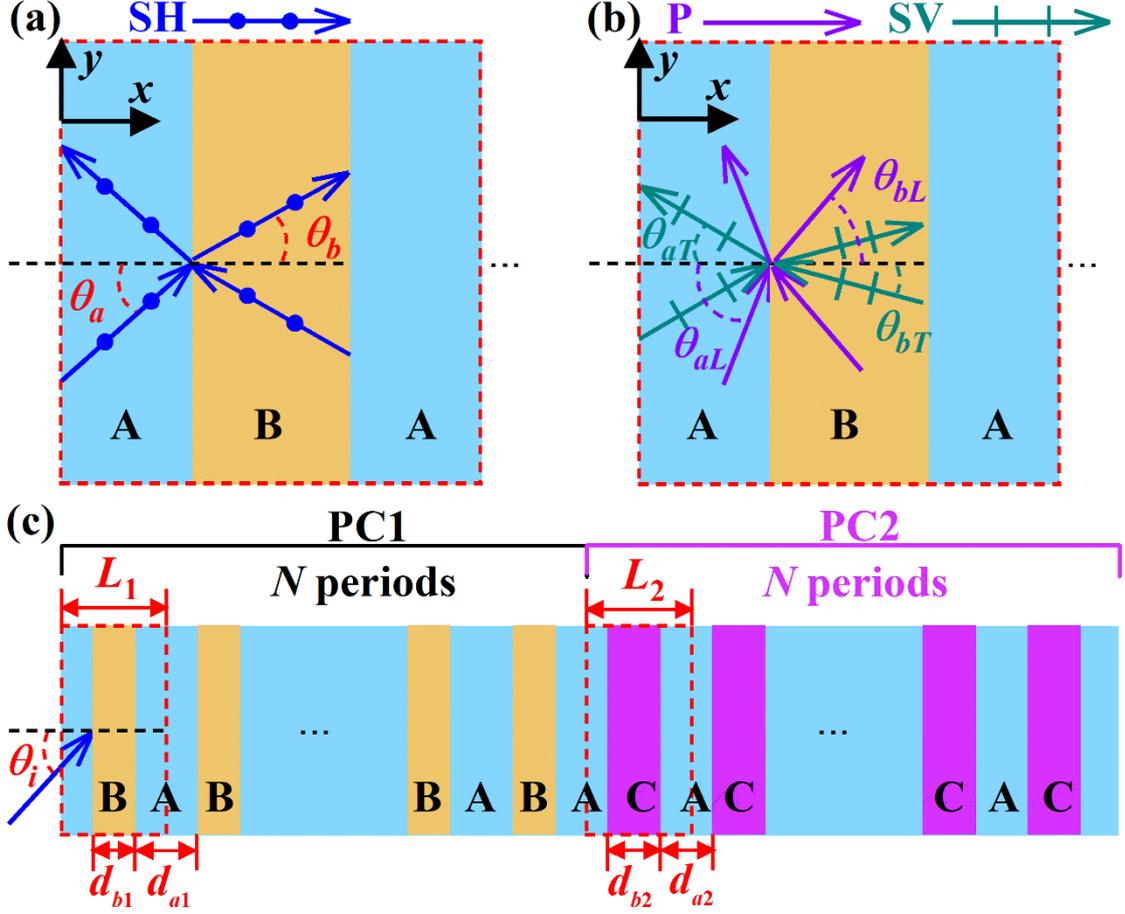

FIG.1. (a) Schematic presentation of the unit cells of 1D PC for the propagation of SH mode. $\theta_a$ and $\theta_b$ are the incident and refracted angles, respectively. (b) Schematic presentation of the unit cells of 1D PC for the propagation of P-SV coupled mode. $\theta_{aL}$, $\theta_{aT}$, $\theta_{bL}$ and $\theta_{bT}$ are the incident angle of P mode, reflected angle of SV mode, refracted angle of P mode and refracted angle of SV mode, respectively. (c) The schematics of the composite structure which is composed of PC1 and PC2. PC1 is composed of materials A and B with thickness $d_{a1}$ and $d_{b1}$, respectively. $L_1 = d_{a1} + d_{b1}$ is the lattice constant of PC1. PC2 is composed of materials A and C with thickness $d_{a2}$ and $d_{b2}$, respectively. $L_2 = d_{a2} + d_{b2}$ is the lattice constant of PC2. $N$ is the period of both PC1 and PC2.

## II. Multiple topological interface states for the SH mode

Firstly, we design a 1D periodic PC to obtain the topological interface states for SH wave. As shown in Fig. 1(a), the unit cell of the PC is composed of the AB layered structures. The transfer matrix (referred to as $M$) for the SH mode is a $2\times 2$ matrix. In this case, the dispersion relation between the Bloch wave number $k$ and the characteristic frequencies $\omega$ is given by

$$\cos(k\Lambda) = \frac{1}{2}(m_{11} + m_{22}). \tag{1}$$

Here $\Lambda$ is the lattice constant and $\Lambda = L_1 = d_{a1} + d_{b1}$ in PC1 and $\Lambda = L_2 = d_{a2} + d_{b2}$ in PC2, respectively. The solved eigenvector of Eq. (1) is $[m_{12}, \exp(iK\Lambda) - m_{11}]^T$, where $m_{11}$, $m_{12}$, $m_{21}$, and $m_{22}$ are the matrix elements of $M$ related to the frequency $\omega$, as shown in Eq. (A8) and Eq. (A9) (see section 1 of supplementary materials). According to Eq. (A8), the band structure of the PC with constant geometric parameters can be regulated by the incident angle $\theta_i$. For the $n$th isolated band, the Zak phase [9,10] is given by

$$\theta_n^{Zak} = \int_{-\pi/\Lambda}^{\pi/\Lambda} \left[ i\int_{\text{unit cell}} dx \cdot \psi_{n,k}^*(x) \frac{\partial \psi_{n,k}(x)}{\partial k} \right] dk. \tag{2}$$

Here $i\int_{\text{unit cell}} dx \cdot \psi_{n,k}^*(x) \partial \psi_{n,k}(x)/\partial k$ is the Berry connection. $\psi_{n,k}(x)$ is the Bloch wave of the displacement field on the $n$th band with wave number $k$, as shown in Eq.(A17)-(A24) (see section 1 of supplementary materials). Besides, the topological property of the $n$th bandgap depends on the summation of Zak phases for all bulk bands below the gap [10]. That can be distinguished by the sign of $\varsigma^{(n)}$, which has the expression

$$\text{sgn}[\varsigma^{(n)}] = (-1)^n \exp(i\sum_{m=0}^{n-1} \theta_m^{Zak}). \tag{3}$$

If two PCs have bandgaps in common frequency region but with different $\text{sgn}[\varsigma^{(n)}]$, the topological interface states will definitely exist, which is the well-known bulk-interface correspondence in 1D system. Thus, through calculating the Zak phase, we can explore the dependency of topological properties on incident angle, and further control the generation and elimination of topological interface states by tuning the incident angle.

To demonstrate this idea, the mass densities and transverse wave velocities of used material are: $\rho_a = 7700 \text{kg/m}^3$ and $c_a = 2400 \text{m/s}$ for cast iron (material A), $\rho_b = 11600 \text{kg/m}^3$ and $c_b = 1133 \text{m/s}$ for lead (material B), $\rho_c = 19500 \text{kg/m}^3$ and $c_c = 1239 \text{m/s}$ for gold (material C). Here, the geometric parameters of PC1 and PC2 are $\Lambda = L_1 = L_2$, $d_{a1} = 0.55\Lambda$, $d_{b1} = 0.45\Lambda$, and $d_{a2} = d_{b2} = 0.5\Lambda$, respectively. Fig.2(a) shows that the two lowest bandgaps I and II (the grey areas) of PC1 undergo an open-close-reopen process with the increase of the incident angle $\theta_i$. At the cross points, corresponding to $\theta_i = 48.2°$, the first bandgap close at the Brillouin zone boundary and the second bandgap close at the Brillouin zone center (see Fig. S2(b) in section 2 of supplementary materials). It is the critical state of the band inversion. For the case of $\theta_i = 18° < 48.2°$, the Zak phases of the lowest three bulk bands, as the red solid lines shown in Fig. 2(b), can be numerically calculated from Eq. (3) to be "$\pi - 0 - \pi$". Consequently, $\text{sgn}[\varsigma^{(n)}]$ for bandgaps I and II are positive and negative ($+$ and $-$), respectively. While for $\theta_i = 72° > 48.2°$, the Zak phases of the lowest three bulk bands change to "$0 - 0 - 0$", as shown in Fig. 2(c), and the corresponding $\text{sgn}[\varsigma^{(n)}]$ for bandgaps I and II become negative and positive ($-$ and $+$), Apparently, $\text{sgn}[\varsigma^{(n)}]$ reverses in both the bandgaps I and II. That means the topological properties of both bandgaps I and II have been changed as the incident angle changes from $\theta_i < 48.2°$ to $\theta_i > 48.2°$.

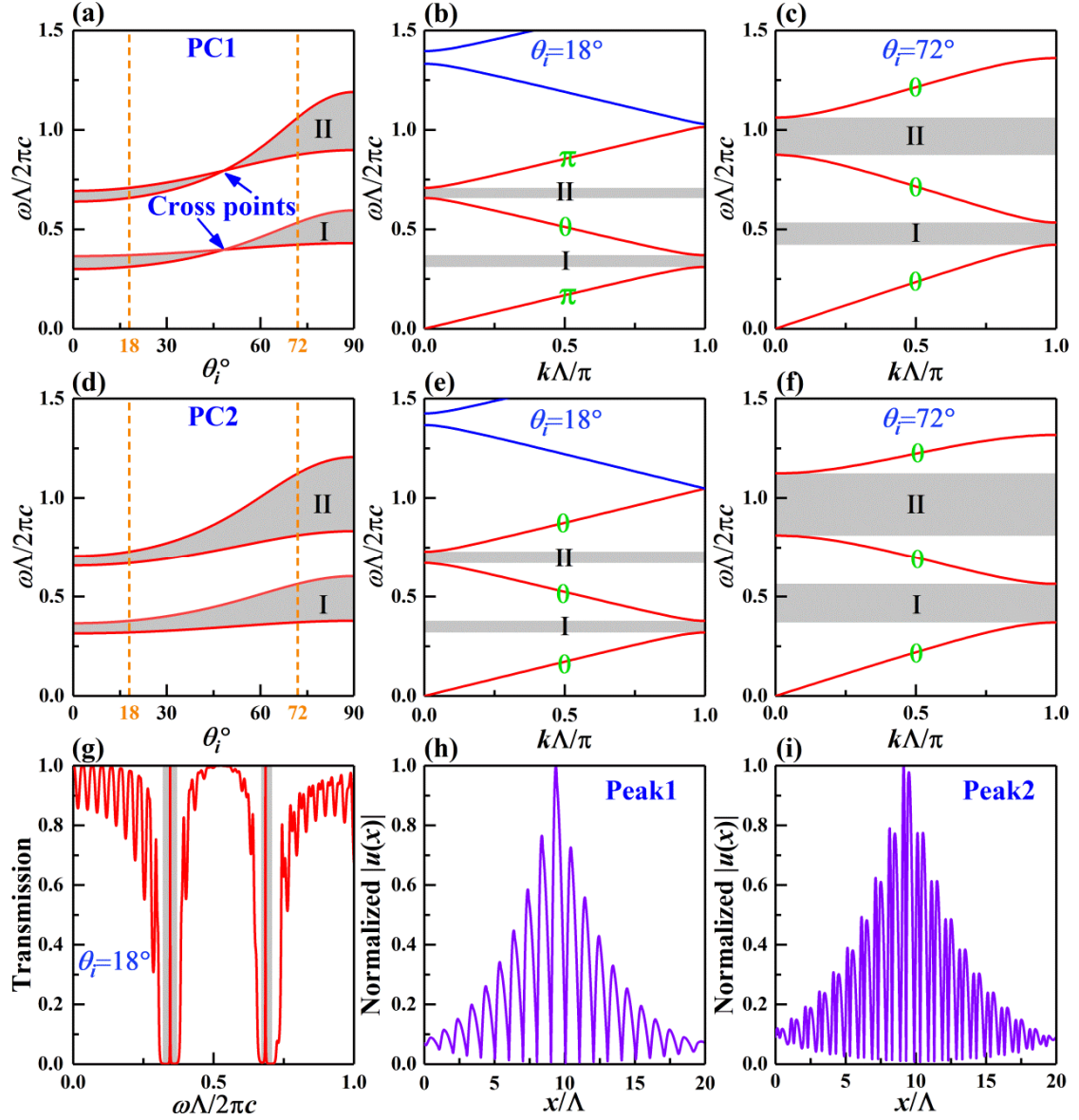

FIG.2. (a) and (d) show the band evolution of PC1 and PC2 versus the incident angle $\theta_i$, respectively. Here, $\omega\Lambda/2\pi c$ is the normalized frequency, and $c = c_a$. (b-c) and (e-f) are the band structures of PC1 and PC2 for incident angle $\theta_i = 18°$ and $\theta_i = 72°$, respectively. And the Zak phases of isolated bands are marked by green number "0" or "$\pi$". The grey areas in (a)-(f) denote the bandgaps. (g) is the transmission spectrum of the composite PC1+PC2 structure at $\theta_i = 18°$. Here, the gray area indicates the common bandgaps of PC1 and PC2. (h) and (i) are the absolute displacement field distributions $|u(x)|$ (normalized by the maximum value) along the x-axis, corresponding to the transmission peaks in (g).

Next for PC2, we also investigate the band evolution and calculate Zak phases at the same incident angles. Fig. 2(d) shows that both the bandgaps I and II of PC2 keep opening as the incident angle $\theta_i$ changing from $0°$ to $90°$. Different from PC1, the Zak phases of the lowest three bulk bands of PC2 remain "$0-0-0$", and then $\text{sgn}[\varsigma^{(n)}]$ for the bandgaps I and II keep negative and positive ($-$ and $+$) as the incident angle changes. Obviously for the PC2, no band inversion happens, and the topological properties of all considered bands remain unchanged. In both bandgaps I and II, $\text{sgn}[\varsigma^{(n)}]$ for PC1 and PC2 are topologically different when $\theta_i < 48.2°$, but topologically identical when $\theta_i > 48.2°$. In addition, whether the topological properties of bandgaps have been changed can be further confirmed by examining the field distribution of the band-edge states [10], which is shown in Fig. S3 in section 2 of supplementary materials.

For both the cases of $\theta_i < 48.2°$ and $\theta_i > 48.2°$, PC1 and PC2 have common bandgaps (see Fig. S2(a) in section 2 of supplementary materials). If PC1 and PC2 are combined to form a composite structure (schematically presented in Fig. 1(c)), according to the analysis of topological properties, topological interface states will certainly exist as $\theta_i$ is smaller than $48.2°$ while disappear when $\theta_i$ increases to greater than $48.2°$. Fig. 2(g) shows the transmission spectrum of the composite structure with $\theta_i = 18°$. Two transmission peaks stem from topological interface states at $\omega\Lambda/2\pi c = 0.346$ and $\omega\Lambda/2\pi c = 0.685$ in the common bandgaps are observed. The corresponding absolute displacement distributions $|u(x)|$ (normalized by the maximum value) along the *x*-axis are shown in Fig. 2(h) and 2(i), respectively. They exhibit the typical field distribution pattern of interface states: the field has maximum intensity on the interface and rapidly decays away from the interface. Fig. S4 in section 2 of supplementary materials presents the transmission peaks versus the period number $N$. The results show that the transmission peaks appear steadily at the same frequencies while have better quality factor with the increase of $N$. Fig. S5 in section 2 of supplementary materials exhibits the evolution of transmission peaks as the increase of incident angle. The topological interface states continuously exist in a wide incident angle ($0° \leq \theta_i \leq 40°$), and they gradually fade away when $\theta_i > 48.2°$.

## III. Topological interface states for the P-SV coupled mode

From the vibration pattern, both P mode and SV mode are the in-plane modes, and they will mutually couple in the case of oblique incidence. Thus, the transfer-matrix of the P-SV coupled system is a $4 \times 4$ matrix $M$, and the detailed derivation can be found in section 3 of supplementary materials. In this case, the characteristic equation of the eigenvalue problem is given by

$$e^{4ik\Lambda} - pe^{3ik\Lambda} + qe^{2ik\Lambda} + re^{ik\Lambda} + 1 = 0. \qquad (4)$$

Here, $e^{ik\Lambda}$ is the eigenvalue related to the Bloch wave number $k$, and $p, q, r$ are the coefficients related to the frequency $\omega$ (see Eq. (B13) in section 3 of supplementary materials). As an unary quartic equation, Eq. (4) has four solutions: $e^{ik\Lambda} = e_1, e_2, e_3$, and $e_4$. When the absolute value of the real part of $e_1$ (or $e_2, e_3, e_4$) is smaller than or equal to one, $k$ is a real number and the corresponding frequencies are in the pass band, otherwise in the bandgap. After obtaining the band structure of the P-SV coupled system, the corresponding Zak phase of the $n$th isolated band can also be calculated by Eq. (2). Similar to the SH system, we can explore the dependency of topological properties on the incident angle, and further control the generation and elimination of topological interface states for the P-SV coupled system by tuning the incident angle. However the band structure is much more complex for the P-SV coupled mode and we carry out the discussion aiming at three different typical bandgaps.

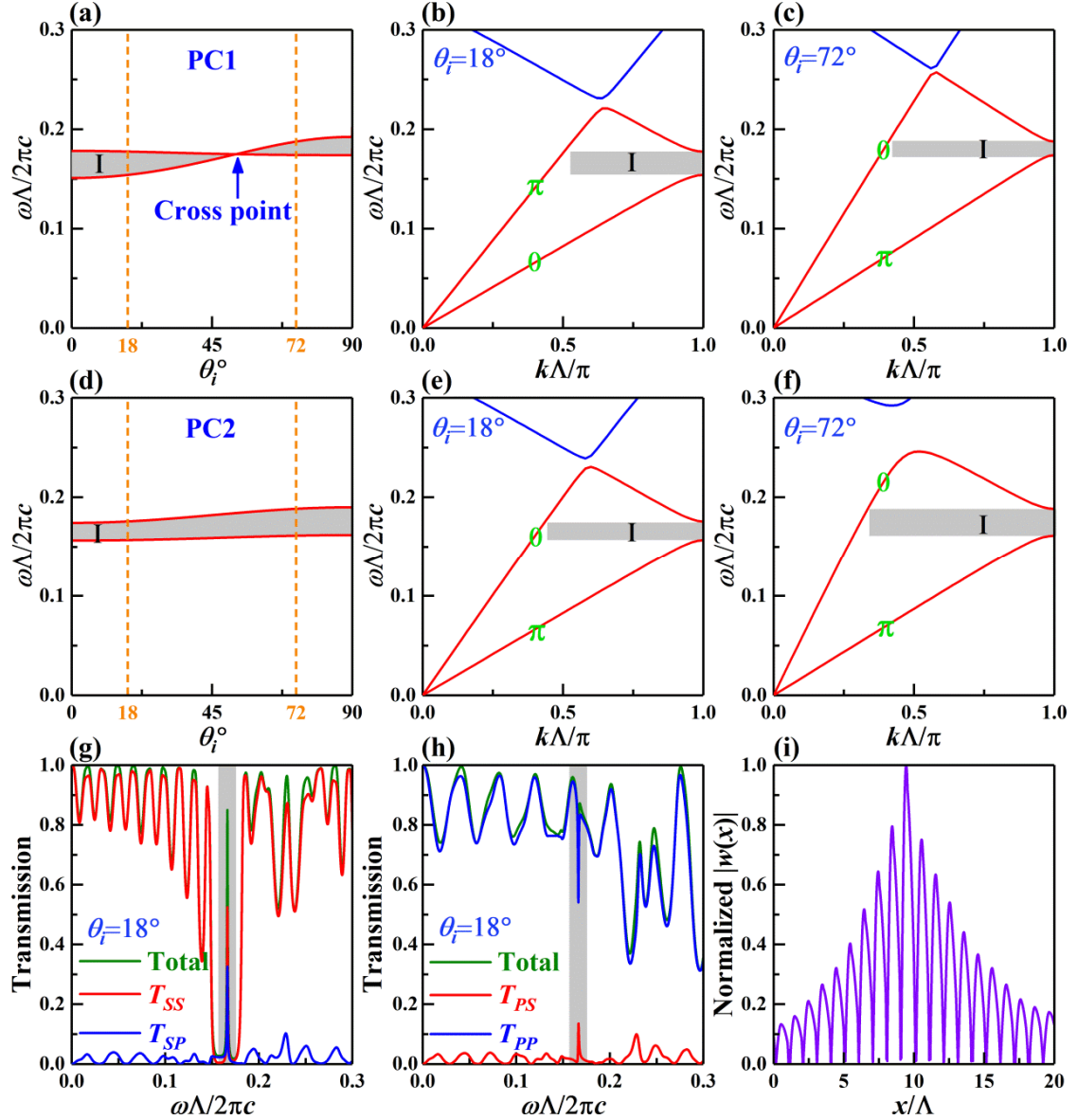

FIG.3 (a) and (d) respectively show the evolution of the partial-polarization bandgap of PC1 and PC2 with the incident angle $\theta_i$. Here, in normalized frequency $c = c_{la}$. (b-c) and (e-f) are the band structures of PC1 and PC2 for incident angle $\theta_i = 18°$ and $\theta_i = 72°$, respectively. The Zak phases of isolated bands are marked by green number "0" or "$\pi$". The grey areas in (a)-(f) denote the bandgaps. (g) and (h) are the transmission spectra of the composite PC1+PC2 structure at $\theta_i = 18°$. $T_{SS}$ and $T_{SP}$ indicate the transmission rate of SV wave and P wave with SV wave incidence, while $T_{PP}$ and $T_{PS}$ indicate the transmission rate of P wave and SV wave with P wave incidence, respectively. Here, the gray area indicates the common bandgaps I of PC1 and PC2. Here $N = 10$. (i) is the normalized total displacement field distributions $|w(x)|$ for the peaks in (g).

**Case 1:**

To demonstrate the topological interface states for this complex P-SV coupled modes, the mass densities, longitudinal wave velocities and transverse wave velocities of chosen materials are: $\rho_a = 7700 \text{kg/m}^3$, $c_{la} = 4500 \text{m/s}$, and $c_{ta} = 2400 \text{m/s}$ for cast iron (material A), $\rho_b = 11600 \text{kg/m}^3$, $c_{lb} = 2490 \text{m/s}$, and $c_{tb} = 1133 \text{m/s}$ for lead (material B), $\rho_c = 19500 \text{kg/m}^3$, $c_{lc} = 3360 \text{m/s}$, and $c_{tc} = 1239 \text{m/s}$ for gold (material C). And the geometric parameters of PC1 and PC2 are $\Lambda = L_1 = L_2$, $d_{a1} = 0.45\Lambda$, $d_{b1} = 0.55\Lambda$, $d_{a2} = 0.35\Lambda$, and $d_{b2} = 0.65\Lambda$, respectively.

First of all, we consider the lowest bandgap. According to the growth trend of dispersion lines, the first bandgap is opened between two SV mode dispersion lines at the Brillouin zone boundary, while this bandgap locates in the first passband of the P mode, which is clearly shown in Figs. 3(b), (c), (e) and (f). That's to say, in the bandgap frequency region, the P wave can pass through the PC but the SV wave cannot (see Fig. S8 in supplementary materials). Then we defined this kind of bandgap as the "partial-polarization" bandgap. Fig. 3(a) shows the evolution of the partial-polarization bandgap of PC1 with the increase of incident angle $\theta_i$. It undergoes an open-close-reopen, i.e. band inversion process. At the cross point, corresponding to $\theta_i = 52.5°$, the bandgap closes at the Brillouin zone boundary (see Fig. S8(b) in section 4.1 of supplementary materials). When $\theta_i$ changes from $18° < 52.5°$ to $72° > 52.5°$, the calculated Zak phases of the lowest two bulk bands switch from "$0 - \pi$" to "$\pi - 0$", as is shown in Figs. 3(b) and 3(c). Consequently, the corresponding $\text{sgn}[\varsigma^{(1)}]$ reverses from negative to positive ($-$ to $+$). Thus, the topological properties of bandgap have been reversed as the incident angle increases from smaller than $52.5°$ to greater than $52.5°$. Different from PC1, Fig. 3(d) shows that the bandgap I of PC2 remains opening with the increase of the incident angle $\theta_i$. The Zak phases of the lowest two bulk bands remain "$\pi - 0$", and then $\text{sgn}[\varsigma^{(\text{I})}]$ keeps positive (+) for all incident angles, as shown in Figs. 3(e) and 3(f). These distinct topological properties of bandgaps can also be confirmed through investigating the displacement field distribution $|u(x)|$ of the band-edge states, as shown in Fig. S9 in section 4.1 of supplementary materials. That means the partial-polarization bandgaps for PC1 and PC2 have different topological properties when $\theta_i < 52.5°$.

Next, we combine PC1 and PC2 to form a composite structure to demonstrate the occurrence of topological interface states. As shown in Fig. 3(g), the transmission peaks of P waves and SV waves resulted from interface states at $\omega \Lambda / 2\pi c = 0.168$ are observed within the common bandgap with SV waves incidence at $\theta_i = 18°$. Some energy has been transferred from SV mode to P mode and the transmittances of SV and P modes are about $0.53$ and $0.32$, respectively. This mode conversion of the interface states changes with different period number $N$ (see Fig. S10 (a)-(d) in section 4.1 of supplementary materials). The normalized total displacement field distribution $|w(x)|$ along the x-axis of the transmission peak is shown in Fig. 3(i). And it also exhibits the typical field distribution pattern of the interface states. What's more, the transmission peaks resulted from topological interface states continuously exist in a wide incident angle ($0° \leq \theta_i \leq 50°$) but disappear when $\theta_i > 50°$, as shown in Fig. S11 in section 4.1 of supplementary materials. For comparison, the transmission spectra with P wave incidence are plotted in Fig. 3(h). It clearly reveals that the waves with the frequencies in partial-polarization bandgaps can penetrate the whole structure because it belongs to pass band for P polarization waves. It should be noted that a small transmission peak belongs to SV mode ($T_{PS} \approx 0.13$) exist in the bandgap. It is also the topological interface state which is transferred from incident P wave. More interesting, as the increase of $N$, the coexistent propagation state and interface state will strongly interacted to generate a Fano resonance. The detailed analysis about the mode conversion can be seen in Fig.S10 in section 4.1 of supplementary materials.

**Case 2:**

Next, we discuss the topological interface states in the omni-polarization bandgap. Here, the mass densities, longitudinal wave velocities and transverse wave velocities of chosen materials are: $\rho_a = 2710 \text{kg/m}^3$, $c_{la} = 6150 \text{m/s}$, and $c_{ta} = 3090 \text{m/s}$ for aluminum (material A), $\rho_{b,c} = 1800 \text{kg/m}^3$, $c_{lb,lc} = 2740 \text{m/s}$, and $c_{tb,tc} = 1600 \text{m/s}$ for epoxy (material B and C). The geometric parameters of PC1 are $\Lambda = L_1$, $d_{a1} = 0.28 L_1$, and $d_{b1} = 0.72 L_1$, while that of PC2 are $L_2 = 0.9 L_1$, $d_{a2} = 0.405 L_2$, and $d_{b2} = 0.595 L_2$.

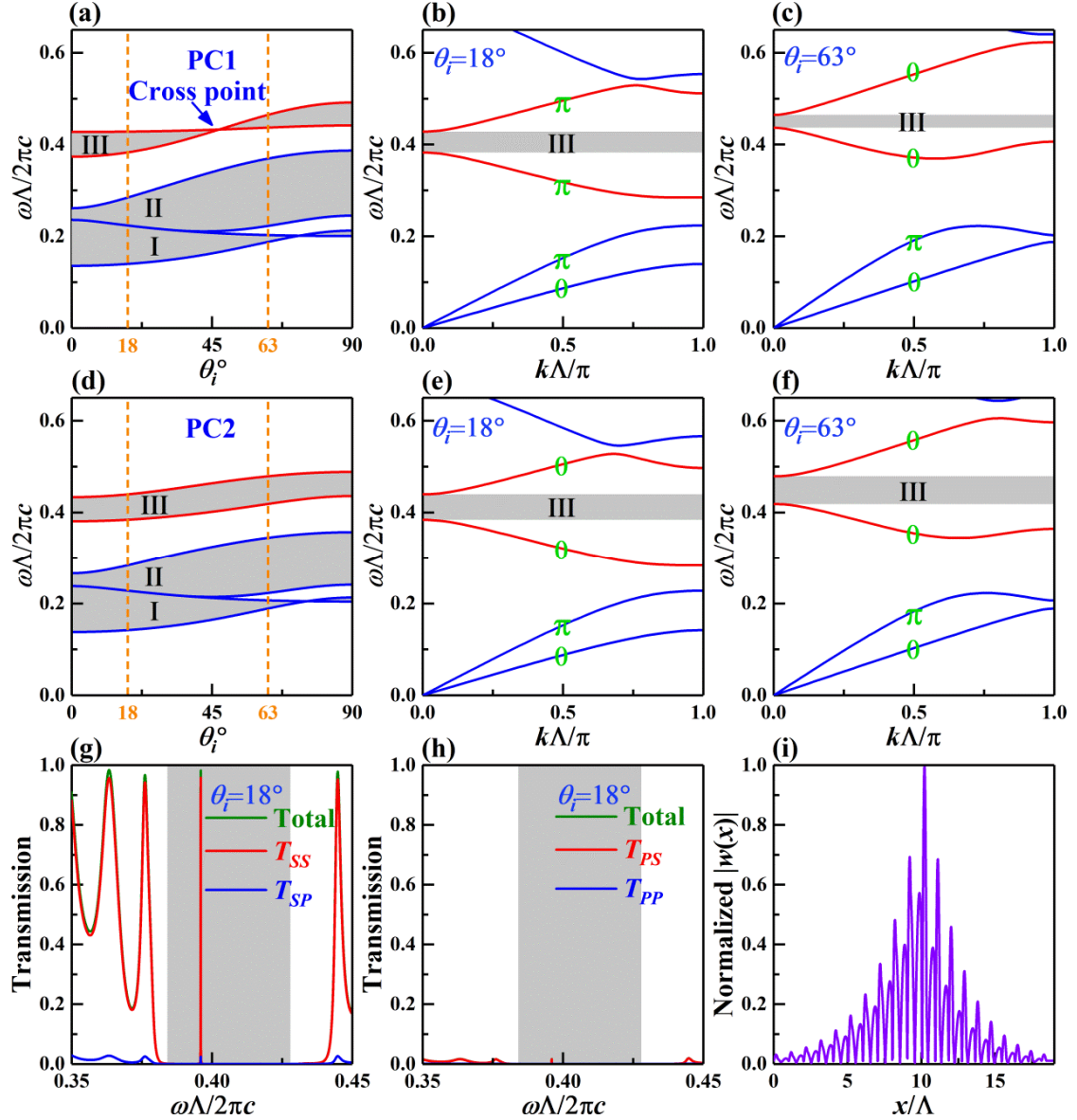

FIG.4. (a) and (d) show the evolution of the omni-polarization bandgaps of PC1 and PC2 with the incident angle $\theta_i$, respectively. Here, in normalized frequency $c = c_{la}$. (b-c) and (e-f) are the band structures of PC1 and PC2 for incident angle $\theta_i = 18°$ and $\theta_i = 63°$, respectively. The Zak phases of isolated bands are marked by green number "0" or "$\pi$". The grey areas in (a)-(f) denote the bandgaps. (g) and (h) are the transmission spectra of the composite PC1+PC2 structure at $\theta_i = 18°$. $T_{SS}$ and $T_{SP}$ indicate the transmission rate of SV wave and P wave with SV wave incidence, while $T_{PP}$ and $T_{PS}$ indicate the transmission rate of P wave and SV wave with P wave incidence, respectively. Here, the gray area indicates the common bandgaps III of PC1 and PC2. Here $N = 10$.

(i) is the normalized total displacement field distributions $|w(x)|$ for the peaks in (g).

Fig. 4(a) shows that bandgap III in high frequencies of PC1 experiences an open-close-reopen process with the increase of the incident angle $\theta_i$. This band inversion process happens between the third and fourth bulk bands. According to the growth trends, although these two bands undergo coupling, the SV mode still dominate (see supplementary materials). Different from case 1, the band inversion located within a wide P mode bandgap. Then we can define the bandgap III as an omni-polarization gap. At the cross point, bandgap closes at the Brillouin zone center with $\theta_i = 47.2°$ (see Fig. S12(b) in section 4.2 of supplementary materials). The band structures of PC1 as well as the Zak phases for $\theta_i = 18° < 47.2°$ and $\theta_i = 63° > 47.2°$ are plotted in Figs. 4(b) and (c), respectively. It clearly exhibits that as the band inversion process occurs, the Zak phases of the lowest two bulk bands remains unchanged "$0-\pi$" while that of the third bulk band flips from "$\pi$" to "$0$". Then $\text{sgn}[\varsigma^{(\text{III})}]$ for changes from negative to positive ($-$ to $+$). But for PC2, the omni-polarization bandgap III remains opening with the increase of the incident angle $\theta_i$ and the topological properties of all bulk bands keep unchanged for $\theta_i = 18°$ and $\theta_i = 63°$, as shown in Figs. 4(d), (e) and (f). Then $\text{sgn}[\varsigma^{(\text{III})}]$ keeps positive ($+$). In a word, the topological properties of the omni-polarization bandgap III are different for PC1 and PC2 when $\theta_i < 47.2°$. Actually, the displacement field distributions of band-edge states for PC1 and PC2 can also support this conclusion (see Fig. S13 in section 4.2 of supplementary materials).

Next, the transmission spectra of composite PC1+PC2 structure are plotted in Figs. 4(g) and (h). Since the band inversion process happens between two SV mode dispersion lines, apparent transmission peak appears in the common omni-polarization bandgap only with SV waves incident at $\theta_i = 18°$. This peak represents the topological interface state, which can be testified by its typical field distribution shown in Fig. 4(i). Different from the topological interface states in partial-polarization, here the topological transition happens in the P mode bandgap. Then very few energy ($T_{SP} \approx 0.02$) can be transferred from SV mode to P mode with SV wave incidence (see Fig. 4(g)). And if we choose P wave incidence, no apparent transmission peaks have been found in this omni-polarization bandgap. More details about the mode conversion can be found in Fig. S14 in section 4.2 of supplementary materials.

**Case 3:**

Next, we discuss the topological properties of bandgap II, which located between bandgap I and bandgap III. Here, the mass densities, longitudinal wave velocities and transverse wave velocities of chosen materials A, B, and C are identical to case 1. And the geometric parameters of PC1 and PC2 are $\Lambda = L_1 = L_2$, $d_{a1} = 0.79\Lambda$, and $d_{b1} = 0.21\Lambda$, $d_{a2} = 0.73\Lambda$, and $d_{b2} = 0.27\Lambda$.

When the first P mode dispersion line meets the second SV mode dispersion line (a folded line on the Brillouin zone boundary), the mutual coupling happens and then a bandgap can be opened, as shown in Figs. 5(b), (c), (e) and (f). This kind of gap is unique for the elastic waves propagating in 1D PC with oblique incidence. It is also an omni-polarization gap but it is opened within the Brillouin zone instead of on the Brillouin zone boundary or center. Fig. 5(a) demonstrates the evolution of the bandgap II of PC1. It also undergoes an open-close-reopen process with the increase of the incident angle $\theta_i$. The bands cross at about $\theta_i = 62.9°$ within the Brillouin zone (see Fig. S16 in section 4.3 of supplementary materials). But different from the band inversion process in case 1 and case 2, here the open-close-reopen process will not bring about any change of Zak phase. When the incident angle changes from $\theta_i = 54°$ to $\theta_i = 72°$, the numerically calculated Zak phases of the lowest three bulk bands remain unchanged as "$\pi - 0 - 0$" and the corresponding $\text{sgn}[\varsigma^{(\text{II})}]$ keeps negative ($-$), as shown in Figs.

5(b) and (c). That means the happening of open-close-reopen process within Brillouin zone will not change Zak phase of bulk bands. The displacement field distributions of band-edge state have also been discussed in Fig. S17 in section 4.3 of supplementary materials. And for PC2, the omni-polarization bandgap II remains opening with the increase of the incident angle $\theta_i$, then the topological properties of all bulk bands remain unchanged. Obviously no topological interface states will exist in the composite PC1+PC2 structure, which result in no transmission peaks in common bandgap as shown in Figs. 5(g) and (h).

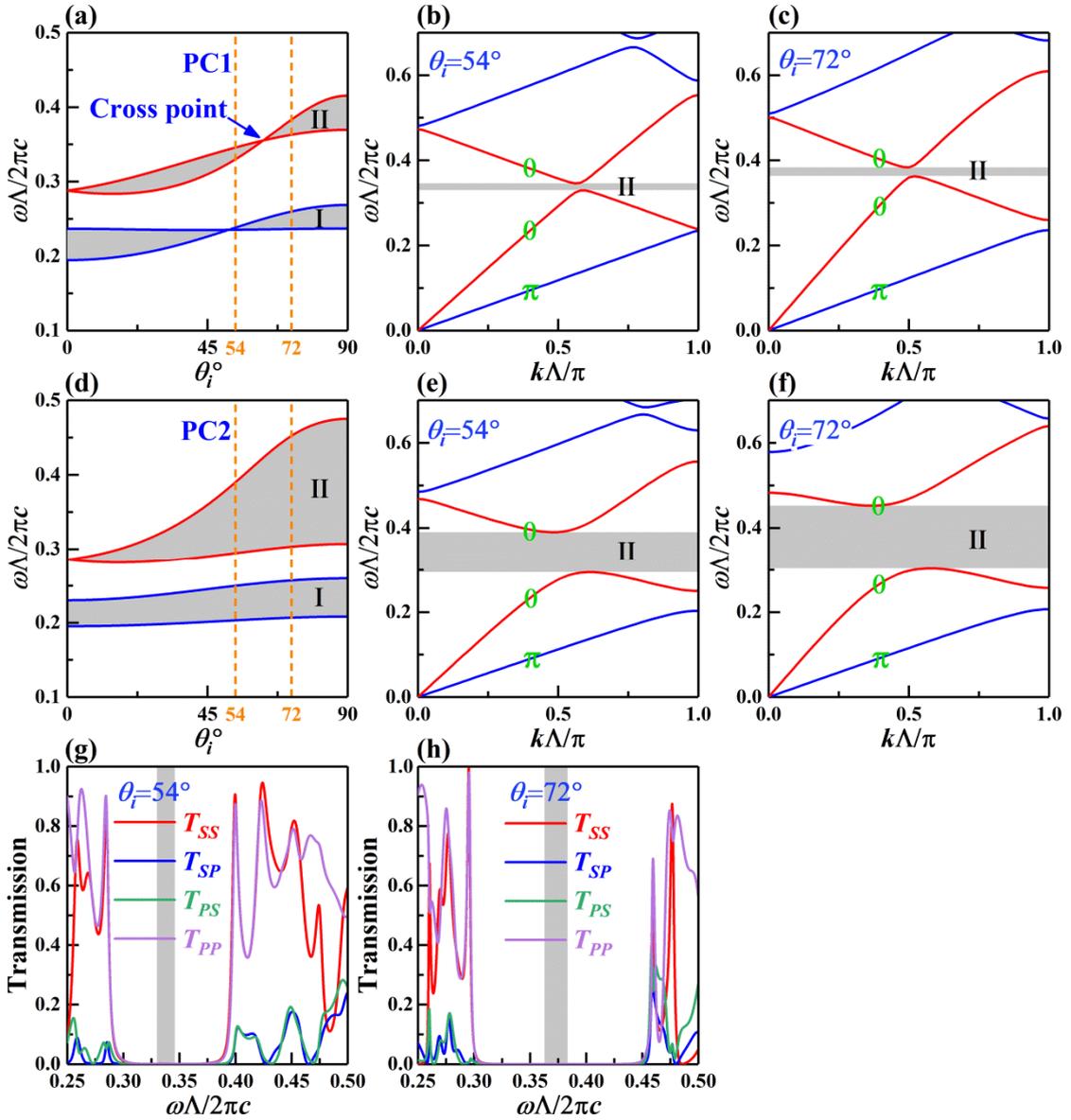

FIG.5. (a) and (d) respectively show the evolution of the omni-polarization bandgap II of PC1 and PC2 with the incident angle $\theta_i$. Here,

in normalized frequency $c = c_{la}$. (b-c) and (e-f) are the band structures of PC1 and PC2 for incident angle $\theta_i = 54°$ and $\theta_i = 72°$, respectively. The Zak phases of isolated bands are marked by green number "0" or "$\pi$". The grey areas in (a)-(f) denote the bandgaps. (g) and (h) are the transmission spectra of the composite PC1+PC2 structure at $\theta_i = 54°$ and $\theta_i = 72°$. The grey areas denote the common bandgaps.

## IV. Conclusion and discussion

In summary, we have systematically demonstrated the topological properties of bulk bands and the existence of the topological interface states for the oblique incidence of elastic waves in PC. For the decoupled SH mode, we achieve multiple topological interface states, which can survive in a wide incident angle region. And for P-SV coupled modes, the topological interface states can exist in both partial-polarization and omni-polarization common bandgaps. The band inversion process happening on the Brillouin zone boundary or center is the necessary condition for the change of topological properties of bulk bands (case 1 and case 2). But if this process happens within the Brillouin zone, the topological properties of bulk bands will not change (case 3). And the band inversion in case 1 and case 2 happens between SV dispersion lines and then the topological of both cases can be excited by SV wave incidence. In the partial-polarization bandgap, some energy of interface states can be transferred from local SV mode to propagating P mode because the topological interface state locates in the pass band of P wave. While in the omni-polarization bandgap, very little energy of interface states can be transferred from SV mode to P mode because the interface state is also in a bandgap of P wave.

In this work we mainly focus on the relation of topological properties and incident angle for the lowest band of PC. Similar results can be obtained in higher frequency bands or more cases may be found in other designed structure, for example the topological interface states in the P mode bandgap but meanwhile located in a SV mode pass band. It should be noted that, For the P-SV coupled system, in low frequencies, the coupling of P mode and SV mode is not very complex and we can still distinguish the dominant polarization of each isolated band. But in high frequencies, the coupling is complex and the polarization mode of band can not be easily distinguished. However the calculation of Zak phase is strict. We can still determine the existence of topological interface states by the analysis of Zak phases. When we investigate the evolution of high frequency bandgaps as the change of incident angle, the band inversion process may happen in the bandgaps below the considered bandgap. Fortunately this band inversion will not change the topological properties of considered bandgap. It may also happen in case 2 and case 3 in this work. For example, in Fig. 5(a) for case 3, the bandgap I also undergoes a band inversion process (the incident angle is about $51.7°$ at the cross point). It will bring about the flip of Zak phases from $0,\pi$ to $\pi,0$ for the lowest two bulk bands. However it will not affect the summation of Zak phases in Eq. (3), and then $\text{sgn}[\varsigma^{(\text{III})}]$ will not change. This result can be extended to bandgaps in more higher frequencies.

## Acknowledgements

We thank Meng Xiao for the helpful discussion. This work was supported by National Natural Science Foundation of China under Number 11874168, the National Key R&D Program of China under Grant 2020YFA0211400, and Open Fund of Guangdong Provincial Key Laboratory of Information Photonics Technology (Guangdong University of Technology), No. GKPT20-06.